\documentclass{elsart}

\usepackage{amssymb}

\def\smileface{\raisebox{-2pt}{$\ddot\smile$}}

\def\deadface{\raisebox{0pt}{${\times\times\atop\frown}$}}

\def\qupobs{|{\rm quantum \ fields;} \ \smileface\rangle}

\def\deadobs{|\deadface\rangle}
\def\qdeadobs{|{\rm quantum \ fields;} \ \deadface\rangle}

\def\tensormult{\otimes} 

\begin{document}

\begin{frontmatter}



\title{New physics
beyond the standard model of particle
physics and parallel universes}

\author{R.Plaga}

\ead{rainer.plaga@gmx.de}

\address{Franzstr.40, 53111 Bonn, Germany}

\begin{abstract}
It is shown that if - and only if - ``parallel universes'' exist, an electroweak vacuum
that is expected to have decayed since the big bang with a high probability might
exist. It would 
neither necessarily 
render our existence unlikely nor could it be observed.
In this special case the observation of certain combinations of
Higgs-boson and top-quark masses -
for which the standard model predicts such a decay - cannot
be interpreted as evidence for new physics 
at low energy scales.  
The question of whether parallel universes exist is of interest to
our understanding of the standard
model of particle physics.
\end{abstract}

\end{frontmatter}
\section{Introduction}
\label{intro}
\newtheorem{thm1}{}
\newtheorem{thm2}{}
A central question of particle physics is whether there is new physics
beyond the standard-model of particle physics\cite{sm} (SM) at energy scales
that can be experimentally reached at the Tevatron or 
the next generation Large-Hadron collider. There is no
completely compelling argument
that would rule out that the standard model is valid up to
extremely high energy scales such as the GUT unification or even
the Planck scale.
\\
One of the best arguments in favour of new physics at moderate
energy scales are experimental indications for a light SM Higgs boson.
When the LEP-2 accelerator at CERN closed down 
finally on November 3, 2000, it had produced slightly more events
that are compatible with being due to the decay of Higgs bosons
with a rest mass of
\begin{equation}
m_{H} = 115 ^{+0.7}_{-0.5} \ {\rm GeV}
\label{higgsd}
\end{equation}
than expected from background
processes\cite{lep56,renton}. 
\\
If the Higgs boson mass, m$_{H}$ and the top
quark mass m$_{t}$ do not satisfy the approximate relationship 
\begin{equation}
m_{t} < 0.43 \ m_{H} + 120.2 \ {\rm GeV}
\label{death}
\end{equation}
and if the SM is correct up to the Planck scale
then the electroweak vacuum is expected to have
decayed from the presently observed ``false'' state
into a ``true'' state since the
beginning of the universe\cite{quiros95}\footnote{Eq.(\ref{death}) 
approximates
an exact calculation given by Espinosa $\&$ Quir\'os\cite{quiros95} 
within its errors for a
strong-interaction constant $\alpha_S$=0.118.}. 
This transition increases all particle masses by many orders
of magnitude and releases a tremendous amount of energy.
These effects make life as we know it impossible in the true vacuum.
Therefore the following conclusion ``C'' is usually drawn:
{\it If the SM is correct up to the Planck scale, 
we conclude from our very existence, that
combinations of m$_t$ and M$_H$ that violate eq.(\ref{death})
are empirically ruled out.}  
\\
The current ``world average'' for the mass of the top quark is\cite{marciano}:
\begin{equation}
m_{t} = 178.0 \pm 4.3 \ {\rm GeV}.
\label{top}
\end{equation}
Eqs.(\ref{higgsd},\ref{top}) are incompatible
with eq.(\ref{death}) on the 1.9-$\sigma$ level\footnote{
A recent preliminary update found
172.7 $\pm$ 2.9 GeV\cite{cdf05}, which corresponds to a
1.1-$\sigma$ discrepancy.} . This disagreement is not compelling
statistically. Moreover, the assumption that the early universe 
went through a very hot phase, 
used by Espinosa $\&$ Quir\'os to derive
eq.(\ref{death}), is not completely certain\cite{isidori}.
Still, it has been convincingly argued by Ellis et al.\cite{jellis},
that the fact that observations favour particle masses
that are ruled out by conclusion C is an argument  
to give up its assumption: the validity of the SM up to very high energies. 
New physics, most likely supersymmetry,
could then stabilize the electroweak vacuum at energies
below 10$^6$ GeV even for the central values of eqs.(\ref{higgsd},\ref{top}).
This argument is of momentous importance for
high-energy physics and deserves to be 
scrutinized from various angles.
\\
Here I argue, that conclusion C might be avoided in a 
completely different way than the one proposed by Ellis et al..
It might fail because parallel universes exist.
As a preparation,
the next section \ref{parallel} reviews the concept
of parallel universes.
Section \ref{parvac} explains why their existence allows to reconcile
a vacuum that decayed with a high probability with our experience
that it didn't. Section \ref{concl} concludes. 
\\
My purpose is only to delineate a special circumstance under which an important argument
about the SM's limits of validity ceases to hold. No anthropic arguments of any
sort are intended.
\\
\section{Parallel universes}
\label{parallel}
``Parallel universes'' are an infinity of distinct universes
that are completely identical to ours 
until a random decision 
makes subsets of them different in the random-decision results.
They might exist as a consequence of various
physical theories. One possibility is that
the universe is spatially infinite and homogeneous
on large scales and contains
infinitely many Hubble-distance sized regions with an identical 
structure because the chance for their formation is finite (parallel universes
separated in Minkowski space)\cite{ellis}. Another option
is that the basic postulates of quantum mechanics can be 
literally extrapolated to human observers, thus leading to
different components after each quantum-mechanical measurement, each
representing a universe with a different measurement result
(parallel universes separated in Hilbert space)\cite{everett,zeh70,deutschb}.
Both possibilities are speculative. 
However, there is general agreement that both are fully compatible
with the current ``concordance model'' of cosmology and standard
quantum mechanics, respectively.
They are no arbitrary additions to these theories
but might be natural consequences of them.
For demonstration purposes I will 
usually assume the latter possibility in the rest of this paper
- Everett's many worlds interpretation of quantum mechanics - without claiming that
this is the only possibility.
\\
Other parallel universes 
do not have any direct influence on ours, either
because they are located far beyond our cosmological horizon
or because rapid
decoherence ensures the absence of any measurable interference effects between
macroscopic universes\cite{joos}.
\section{Parallel universes and vacuum decay}
\label{parvac}
\subsection{Existence after vacuum decay?}
Formally standard quantum theory yields the following
state of the quantum-mechanical system: ``unstable vacuum and quantum fields''
at the beginning of the universe\footnote{The present discussion greatly
simplifies what possibly was a complex dynamical process involving many
different vacuum states.}:
\begin{eqnarray}
|\Psi_{initial}\rangle =  |{\rm false \ vacuum}\rangle 
\tensormult |{\rm quantum \ fields}\rangle
\label{vacu_ini}
\end{eqnarray}
Here $|{\rm false \ vacuum}\rangle$ represents the usual quantum-mechanical false-vacuum state,
and ${\rm |quantum \ fields}\rangle$ represent the quantum fields in nature, like the quark, electron
and photon fields.
If the vacuum is unstable, unitary evolution due
to a standard-model operator U$_{SM}$ has
evolved this state within the 13.6 billion years since the origin of the
universe into\footnote{Relative-phase angles between components
of a state are
unimportant for the present purpose. They have been set to 0 throughout this
paper.}:
\begin{eqnarray}
|\Psi_{current}\rangle =  U_{SM} |\Psi_{initial}\rangle = 
\nonumber
\\
\sqrt{1-P_{decay}} |{\rm false \ vacuum}\rangle 
\tensormult\qupobs+ 
\nonumber
\\
\sqrt{P_{decay}} |{\rm true \ vacuum}\rangle \tensormult\qdeadobs
\label{vacu}
\end{eqnarray}
$|{\rm true \ vacuum}\rangle$ is the new true-vacuum state into which the false-vacuum state decays.
$\qupobs$ symbolizes the state of the quantum fields entangled with a false vacuum
that now form (besides many other things) our 
humanity. $\qdeadobs$ represent quantum fields
entangled with a true vacuum, $\deadobs$ symbolizes that these fields cannot form a living
humanity.
P$_{decay}$ is the probability that vacuum decay has occurred 
up to to a given moment in time.
If the vacuum decays, P$_{decay}$ increases with time 
and can be extremely near to 1 today.
However, the decay remains exponentially suppressed, i.e.
1 -- P$_{decay}$ remains finite for all reasonable parameters of
the SM. E.g. for the central mass values in eqs.(\ref{higgsd},\ref{top}) one
obtains a
decay probability\footnote{
The following numbers are taken from the work of Isidori, Ridolfi $\&$ Strumia\cite{isidori}
who calculated the decay probability under the conservative assumption that
the universe was always at zero temperature. The effect of 
higher temperatures in the early universe is to push 1-P$_{decay}$
even closer to 0.} up to the present time of P$_{decay}$ $\approx$  1 - e$^{-75}$ but
the decay is still exponentially suppressed by the tiny factor of e$^{-404}$.
\\
It is conventional wisdom that only one of the two components in 
eq.(\ref{vacu}) exists with a probability given by the Born rule  
(e.g. because one of them vanishes in a ``collapse of the wave function''). This
assumption predicts that with a probability P$_{decay}$ the present
state of the ``vacuum-humanity'' system is:
\begin{equation}
|\Psi_{current}\rangle = 
 |{\rm true \ vacuum}\rangle \tensormult\qdeadobs.
\end{equation}
Because P$_{decay}$ is very near 1 if the vacuum is unstable, it is usually
concluded that the vacuum must be at least meta-stable (i.e. stable
on time scale longer than the age of the universe, so that P$_{decay}$ $\ll$ 1
today).
Conclusion C then follows as a corollary. 
\\
Alternatively - if parallel worlds exist e.g. 
in Everett's many worlds interpretation - both components of eq.(\ref{vacu}) continue to
coexist (thus forming parallel universes), 
so its first component ``$|{\rm false \ vacuum}\rangle$ 
$\tensormult \qupobs$'' continues to exist with probability=1,
no matter how small 1-P$_{decay}$ becomes. 
In other words:
because the overall quantum-mechanical amplitude of our universe is not measurable for us,
all we can safely conclude from the empirical fact of our existence is that
this first component did not completely vanish up to now.
Human consciousness is not fully understood, yet, so it might remain
controversial if the state described by eq.(\ref{vacu}) is
really compatible with human experience if (1-P$_{decay}$) $\ll$ 1. However,
all that is needed to draw this papers's conclusion is the
undeniable fact that  
such a compatibility cannot be ruled out, presently.
The proposition of compatibility can be formulated as
the following complement to the Born rule:
\begin{thm1}
If parallel universes do exist
the Born rule must be applied only to those state components
that contain observers that continue to exist after the measurement.
\label{theo1}
\end{thm1}
\subsection{Detecting vacuum decay?}
\label{detvac}
If P$_{decay}$ is near one, with a very
high probability many transition events from
the false to the true electroweak vacuum 
have taken place on the future light cone of a given 
observer.
Should the observer notice anything of these events?
\\
Vacuum-transition events proceed via nucleation of a bubble
that expands and eventually converts the
whole universe into the true-vacuum phase.
While the probability P$_{decay}$ for bubble formation does not
depend strongly on the uncertain depth and properties of the ``true-vacuum'',
the propagation speed of the ``bubble wall'' does.
In the likely case that the 
transition proceeds as a detonation wave through the false
vacuum, the bubble wall moves with a speed v=(1-k)c where
k $\approx$ 0.1/$\alpha$ for $\alpha$ $\gg$ 1. $\alpha$ is
the ratio of the energy density in the true vacuum to the
much smaller one of the false vacuum\cite{steinhardt}. In the standard model
the true vacuum is unbounded from below (i.e. $\alpha$ = $\infty$), so
the propagation proceeds formally with exactly c.
If the effective potential receives
a large positive contribution from quantum gravity near the
Planck-energy scale, k $\approx$ 10$^{-121}$\cite{quiros95},
i.e. the bubble
wall still moves with a speed that is extremely close to one of light. This
makes each transition unobservable for a human being. 
Relativistic causality prevents observers 
within the same component of the state function
as the nucleation event to see any signal from the transition
earlier than $\Delta$t = d $\times$ k/c.
The time scale $\Delta$t is much shorter than the one of
human consciousness $\Delta$t$_{hc}$ $\approx$ msec,
even for transitions that took place at the farest possible
distance d of about 10$^{10}$ light years. 
After the detonation front 
has passed such observers no longer exist.
This makes it impossible for a human to consciously 
register any of the transitions. 
\\
Observers on a different component of the state function
are also not affected by the transition:
the `` peaceful coexistence'' of vacua implied by eq.(\ref{vacu}) in
the many-worlds interpretation
is compatible with experience because rapid
decoherence ensures the absence of any measurable interference effects between
the components\cite{joos}. The very fact that parallel universes different
from ours do not directly
influence us, ensures that they can even be  
in the dreaded ``true vacuum'' state. 
\\
Summarizing, if parallel universes exist
we cannot rule out with certainty that we live in a vacuum
that decays on a short time scale. No definite conclusions
based on vacuum stability
about the validity of the SM can then be drawn.
\section{Conclusion}
\label{concl}
The question whether parallel universes exist 
has been shown to be of interest to
our understanding of the standard
model of particle physics. 
If its answer is affirmative
a light Higgs boson leading to a rapidly decaying electroweak
vacuum might be compatible with experience.
On the one hand this possibility might be interpreted as a slight damper to our hope to
find experimental evidence for new physics at LHC or even the Tevatron.
On the other hand, the prospect that we
live in a rapidly decaying electroweak
vacuum seems worthy of further
investigation.

\section*{Acknowledgements}
Discussions with the late James Higgo were important
to form the basic idea for this manuscript.
Kari Enqvist and Gino Isidori patiently answered many
questions concerning vacuum instability.
Erich Joos and Alvaro de Rujula constructively 
criticised earlier versions of this manuscript.
An anonymous referee raised an important
point and helped to improve the manuscript.
I thank them all!



\end{document}